\documentclass[12pt]{article}
\begin{document}
\begin{center}
{\large Does a Dynamical System Lose Energy } \\
{\large by Emitting Gravitational Waves?}
\\
F. I. Cooperstock
 \\{\small \it Department of Physics and Astronomy, University 
of Victoria} \\
{\small \it P.O. Box 3055, Victoria, B.C. V8W 3P6 (Canada)}
\end{center}
\begin{abstract}
We note that  Eddington's  radiation damping calculation of a spinning rod fails to account
for the complete mass integral as given by Tolman.  The missing stress contributions
precisely cancel the standard rate given by the 'quadrupole formula'.  This indicates that 
while the usual 'kinetic' term can properly account for dynamical changes in the source, the
actual  mass is conserved.  Hence gravity waves are not carriers of energy in vacuum.  This
supports the hypothesis that energy including the gravitational contribution is confined to 
regions of non-vanishing energy-momentum tensor  $T_{ik}$. 
\end{abstract}
%\vspace*{1truecm} 
%%%%%%%%%%%%%%%%%%%%%%%%%%%%%%%%%%%%%%%%%%%%%%%%%%%%%%%%%
.
\\
\end{document}